\documentclass[12pt,a4paper]{article}

\usepackage{amscd,amsfonts,amsmath,amssymb,amsthm,bbm,bm,latexsym,mathrsfs}
\usepackage{epsfig,graphics,graphicx,subfigure,longtable,natbib}
\usepackage[english]{babel}
\usepackage[usenames]{color}
\usepackage{bookmark}
\usepackage{booktabs}
\usepackage{sgame}
\usepackage[top=1.7in, bottom=1.7in, left=0.5in, right=0.5in]{geometry}
\allowdisplaybreaks
\usepackage{caption}
\usepackage{epstopdf}
\usepackage{float}
\usepackage{enumerate}
\usepackage{url}
\usepackage{rotating}
\usepackage{dcolumn}
\usepackage{colortbl}
\usepackage{lscape}
\usepackage[flushleft]{threeparttable}

\newcolumntype{d}[1]{D{.}{.}{#1}}
\makeatother

\theoremstyle{remark}

\theoremstyle{plain}

\begin{document}

\title{\vspace{-50pt}{Comparing the forecasting of cryptocurrencies by Bayesian time-varying volatility models}} 

\author{
\hspace{-20pt} 
Rick Bohte and 
 Luca Rossini\thanks{Address: School of Business and Economics. Department of Econometrics and Operations Research, Vrije Universiteit Amsterdam, De Boelelaan 1105, 1081 HV Amsterdam, The Netherlands. \protect\\ Corresponding Author: Luca Rossini, \href{mailto:luca.rossini87@gmail.com}{luca.rossini87@gmail.com}.} 
        \\ 
        {\centering {\small Vrije Universiteit Amsterdam, The Netherlands}}}

\date{}
\maketitle

\abstract{
This paper studies the forecasting ability of cryptocurrency time series. This study is about the four most capitalized cryptocurrencies: Bitcoin, Ethereum, Litecoin and Ripple. Different Bayesian models are compared, including models with constant and time-varying volatility, such as stochastic volatility and GARCH. Moreover, some crypto-predictors are included in the analysis, such as S\&P 500 and Nikkei 225. In this paper the results show that stochastic volatility is significantly outperforming the benchmark of VAR in both point and density forecasting. Using a different type of distribution, for the errors of the stochastic volatility the student-t distribution came out to be outperforming the standard normal approach.\\

}

%
\section{Introduction}\label{s:introduction}
Nowadays it is more common to handle your affairs online. According to the World Payments Report (Capgemini, 2017), the electronic payments are expected to increase by almost 11\% each year worldwide from 2015 to 2020. The world is becoming more online accessible due to innovations and modern technology. Online investing on the open market is due to technology much easier to do, for example there are applications like eToro, Robinhood and Plus500 where people can invest money with their mobile devices. 

In the last decades a new type of currency is launched on the financial market and has gained importance. In particular, it is a virtual currency of which the main feature is the total absence of any intrinsic value. In 2009, Nakamoto documented the creation of the first decentralized cryptocurrency, called Bitcoin. Since its introduction, it has been gaining more attention from the media, the finance industry, and academics. There are several reasons for this interest, firstly Japan and South Korea have recognized Bitcoin as a legal method of payment (Bloomberg, 2017a; Cointelegraph, 2017). Second, some central banks are exploring the use of cryptocurrencies (Bloomberg, 2017b). Third, the Enterprise Ethereum Alliance was created by a large number of companies and banks to make use of cryptocurrencies and the related technology called blockchain (Forbes, 2017). These were just three of the many reasons why this interest in cryptocurrencies spiked. After the introduction of Bitcoin, a large variety of more cryptocurrencies (around 1000) were created and became a new investment opportunity for trades. Hereafter, a short overlook of the four most important cryptocurrencies is described.

Bitcoin (BTC) is based on decentralization which means that it is controlled and owned by its users. This decentralization is often criticized due to the lack of control over the whole system. Despite this criticism, Bitcoin increased in value from a couple of cents in the beginning (2009) to about 20,000 US dollar at the end of 2017. Ethereum (ETH) is also decentralized and features smart contract functionality. Due to this contractual agreement there is no possibility of fraud, downtime, third party interference or censorship. The researcher and programmer (Vitalik Buterin) proposed it in late 2013 and Ethereum went live at the end of July in 2015. 

Ripple (XRP) is founded by Ryan Fugger in 2004. It is a Blockchain network that incorporates both a currency system known as XRP and a payment system. This enables real-time international payments and is therefore currently used by multiple banks. Litecoin (LTC) was created in 2011 by Charles Lee and is based on the same peer to peer protocol used by Bitcoin. It is often considered Bitcoin's rival due to its improvements in transactions, these transactions are significantly faster than Bitcoin. Therefore it could be particularly attractive in certain situations to invest in.

Recently, researchers have started to study cryptocurrencies by applying different models and techniques. However, apart from Catania et al. (2019), a forecasting analysis of cryptocurrencies has not been strongly used and proposed. This paper tries to continue the analysis initialised by Catania et al. (2019) and to improve it by comparing different multivariate models for point and density forecasting of the four most capitalized cryptocurrencies previously described.

In order to study and forecast the cryptocurrencies, vector autoregressive models and moreover its extension to time-varying volatility have been introduced. Vector autoregressions (VARs) are used in models for empirical macroeconomic applications. VARs are introduced by Sims (1980) and have been widely adopted for forecasting and analysis of macroeconomic variables. The formulation of VARs is simple, however they tend to forecast well and are often used as the benchmark to compare the performance of forecasts among models. Sims and Zha (2006) have emphasized the value of volatility modeling for improving efficiency. Accordingly, taking time variation in volatility into account should improve the estimation of a  VAR-based model and inference common in analysis of macroeconomic variables. Modeling changes in volatility of VARs should also improve the accuracy of density forecasts. Forecast densities are potentially either far too wide or too narrow, due to shifts in volatility. D’Agostino et al. (2013) show that the combination of time-varying parameters and stochastic volatility improves the accuracy of point and density forecasts. One application of these regressions on a macroeconomic level is investing in assets, stocks and, as a purpose for this paper, in cryptocurrencies as mentioned above. 

VAR models could have many parameters if they include many lags, however using non-data information and turning it into priors is found to be greatly improving the forecast performance. In Bayesian estimation algorithms, the stochastic volatility specification is computationally tractable, while in frequentist estimation it is captured with a single model. This is one of the reasons why in this paper the Bayesian approach is used. Another reason is that the Bayesian approach gives some advantages in parameter uncertainty, computing of probabilistic statements and estimation with many parameters. As a standard procedure the normal distribution is often used as a distribution of the so called "noise". For this research not only the normal distribution, but also the student-t distribution is used for modeling the errors. 

A strong improvement of our paper is the introduction of time-varying specifications for multivariate models for better forecasting the cryptocurrencies behavior. In particular, the use of time-varying volatility jointly with the multivariate time series is of interest for capture the possible heteroscedasticity of the shocks and non-linearities in the simultaneous relations among the different cryptocurrencies in the models. Moreover, taking into account the time variation in volatility improves the VAR-based estimation and inference that have been showed in the preliminary cryptocurrencies analysis.

Our results show that including time-varying volatility and in particular stochastic volatility provides forecasting gains in terms of point and density forecasting relative to the multivariate autoregressive model. The inclusion of cryptopredictors can lead to better forecasting with respect to the benchmark but not strong improvements with respect to time-varying volatility models with only lags of the cryptocurrencies included. Directional predictability indicates that using stochastic volatility with heavy tails can be used to create profitable investment strategies.

The content of this paper is structured as follows: in Section 2 some literature which is used as research background is being reviewed, especially research in the field of Bayesian VARs and cryptocurrencies. Section 3 describes the data. Section 4 presents our models, estimation methodology and metrics used to assess our results, which are discussed in Section 5 together
with the major findings. Finally, Section 6 concludes.

\section{Literature review}
Cryptocurrencies is becoming a hot topic in academia and outside of it. In particular, in last years, the interest in cryptocurrencies has exploded from around 19 billion in February 2018 to around 800 billion in December 2017, so a lot of research has been done about this subject. Although Bitcoin is a relatively new currency, there have already been some studies on this topic. 

Hencic and Gourieroux (2014) investigated the presence of bubbles in the Bitcoin/US Dollar exchange rate by applying a non-causal AR model, the dynamics of the daily Bitcoin/USD exchange rate shows episodes of local trends, which can be modelled and interpreted as speculative bubbles. Cheah and Fry (2015) focused on the same issue, as with many asset classes they show that Bitcoin exhibits bubbles. They find empirical evidence that the fundamental price of Bitcoin is zero. The volatility of six major currencies against the volatility of the Bitcoin was measured by Sapuric and Kokkinaki (2014), the results indicate a high volatility for the Bitcoin exchange rate. Then Chu, Nadarajah, and Chan (2015) did a statistical analysis of the log-returns of the exchange rate of the Bitcoin against the US Dollar and the generalized hyperbolic distribution is shown to give the best fit. Yermack (2015) wondered whether the Bitcoin can be considered a real currency on the financial market. 

Fernández-Villaverde and Sanches (2016) analysed privately issued fiat currencies and checked the existence of price equilibria and show that there exists an equilibrium in which price stability is consistent with competing private monies. However they also conclude that the value of private currencies monotonically converges to zero by equilibrium trajectories. Dyhrberg (2016) shows that the movements of the volatility of the Bitcoin has several similarities to gold and the dollar. Bianchi (2018) investigated if there is a relationship between returns on cryptocurrencies and traditional asset classes. There was a mild correlation with some commodities, but not that many macroeconomic variables. 

Catania, Grassi and Ravazzolo (2018) showed that predicting volatility can be improved by using leverage and time-varying skewness at different forecast horizons. Hotz-Behofsits, Huber and Zorner (2018) used time-varying parameter VAR with $t$-distributed measurement errors and stochastic volatility to model three cryptocurrencies: Bitcoin, Ethereum and Litecoin. Griffin and Shams (2018) investigated whether a cryptocurrency called: Tether, is directly manipulating the price of Bitcoin, increasing its predictability. By using algorithms to analyze the data they find that purchases with Tether go along with sizable increases in Bitcoin prices.

In 2019, there are more studies done on cryptocurrencies. Muglia, Santabarbara and Grassi (2019) investigated the predictability of the S\&P 500 by the movement of the Bitcoin, they show that Bitcoin does not have any direct impact on the predictability of the S\&P 500. Catania, Grassi and Ravazzolo (2019) found that point forecasting is statistically significant for the Bitcoin and Ethereum when using combinations of univariate models. They also conclude that density forecasting for all four cryptocurrencies is significant when relying on time-varying multivariate models.

The exercise in this paper is generalised to multivariate models where the four cryptocurrencies are predicted jointly using Bayesian VAR models with stochastic volatility as in Koop and Korobilis (2013). Johannes, Korteweg, and Polson (2014) predicted stock prices using time-varying parameter and stochastic volatility VAR models and find statistically and economically significant portfolio benefits for an investor who uses models of return predictability. 

Many institutions tried to investigate the relationship between the Bitcoin and the stock market. In an article by Bloomberg (2018), analysed stated that "big investors may be dragging Bitcoin toward Market correlation", investors looking for high gains may be attracted to the increasing risk of this cryptocurrency. Stavroyiannis and Babalos (2019) studied the relation between the Bitcoin and the S\&P500 and found that it does not hold any of the hedge, diversifier, or safe-haven properties and the intrinsic value is not related to US markets. 

There are still no studies that can confirm that Bitcoin is a good stock market predictor. This paper tries to fill the gap, analyzing whether the Bitcoin, Ethereum, Litecoin and Ripple can be forecasted by its lags and other macroeconomic variables.

\section{Data}\label{s:data}
\begin{figure}[h!]
  \centering
  \includegraphics[width=1\linewidth]{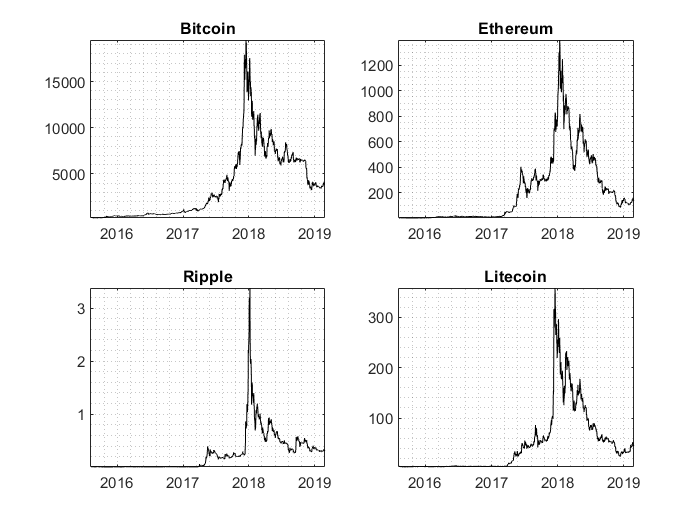}
  \caption{Price of the four cryptocurrencies from August 8, 2015 till February 28, 2019.}
  \label{fig:notrans}
\end{figure}

The data collected for the sample spans from August 8, 2015 till February 28, 2019, giving a total of 1301 observations. The data can be seen in Figure \ref{fig:notrans}, it shows a big spike around the end of 2017. China’s “Big Three” exchanges were pending closure around that time, however the cryptocurrencies were largely buoyed by a bullish sentiment and went up. At December 2017 the peaks were reached and a couple days later they dropped. At this time cryptocurrencies are mainly considered as an alternative investment, due to the fact that their use for payment is still limited. This can create correlations with other assets in the financial market for at least two main reasons. The first regards investors, who usually allocate wealth in a global portfolio and hedge across investments; the second relates to market sentiments that spread fast among different assets. See Bianchi (2018) for similar arguments.

\begin{figure}[h!]
  \centering
    \includegraphics[width=\linewidth]{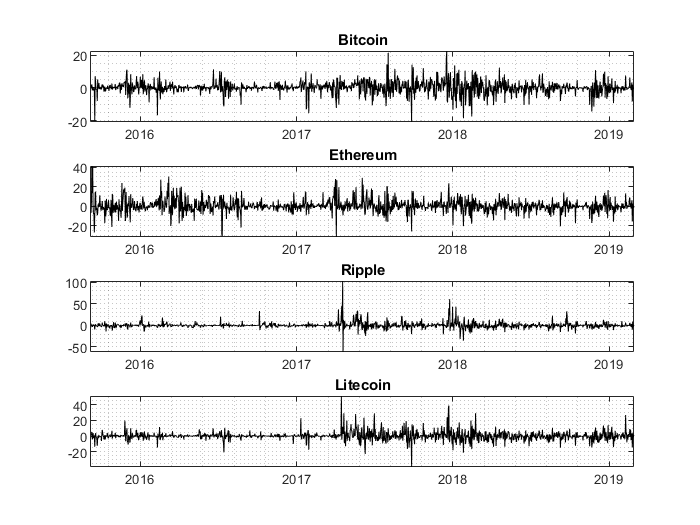}
    \captionof{figure}{Daily log returns of the four cryptocurrencies.}
    \label{fig:daily1}
\end{figure}

In this paper, we have considered different cryptopredictors as described below. The choice of these cryptopredictors is due to the fact that possible correlations between cryptocurrencies and these assets can be created, because Bitcoin and other currencies are considered as an alternative investment and their use as payment is still poor. 
We use the following list of predictors for cryptocurrencies as stated in Catania et al. (2019) as proxying market sentiments: international stock index prices (the S\&P 500, Nikkei 225 and Stoxx Europe 600); commodity prices (gold and silver); interest rates (the 1-month and 10-year US Treasury rates); and the VIX closing price. 
In order to study the possible dependence between cryptocurrencies, a transformation is necessary. The percentage daily log returns of cryptocurrencies will be computed as follows:

\begin{equation*}
    y_t = 100 * \log(S_t/S_{t-1}), 
\end{equation*}
where $S_t$ is the price on day $t$ and $y_t$ is the cryptocurrency log return. Table \ref{table:statistics} reports the descriptive statistics of the cryptocurrencies. In Figure \ref{fig:daily1} the transformed data is plotted against time, as documented in Chu et al. (2015), the cryptocurrencies display high volatility, non-zero skewness, very high kurtosis and several spikes.

\begin{table}[h!]
\centering
\caption{Descriptive statistics, calculated between 08/08/2015 and 02/28/2019}
\begin{tabular}{l r r r r}
\hline
Coin & Bitcoin & Ethereum & Ripple & Litecoin \\ \hline
Maximum & 22.5119 & 41.2337 & 102.7356 & 51.0348\\
Minimum & -20.7530 & -31.5469 & -61.6273 & -39.5151\\
Mean & 0.2071 & 0.4001 & 0.2781 & 0.1912 \\
Median & 0.2343 & -0.0884 & -0.3537 & 0.0000 \\
Std Dev. & 3.9543 & 6.7950 & 7.4433 & 5.7424 \\
Skewness & -0.2624 & 0.4898 & 3.0179 & 1.2631 \\
Kurtosis & 7.8178 & 7.6368 & 42.6234 & 15.3417 \\ \hline
\label{table:statistics}
\end{tabular}
\end{table}

The Ripple has the highest volatility due to the highest kurtosis. The Litecoin has also a high volatility but not that high compared to the Ripple. The other two (Bitcoin and Ethereum) are compared to the aforementioned cryptocurrencies less volatile, however the kurtosis is still far away from the normal distribution, which has a kurtosis of three. Another interesting statistic is the skewness, the Bitcoin is the only one with a negative skewness. This indicates that the tail is at the left side of the distribution, so the probability of lower values than the mean is higher than the normal distribution, which has a skewness of zero. With a positive skewness, this is the case for the other cryptocurrencies, the opposite is true. As before, the Ripple has the highest skewness, which indicates that the Ripple has the highest probabilities of higher values than its mean. 

In Figure \ref{fig:daily1} the transformation of daily log returns is shown. This gives some more insight in the cryptocurrencies. The Ripple is the crypto which is the most volatile which the descriptive statistics also indicated. Also Ethereum stands out in the first half and after that it is more stable, which means that it is less volatile. Bitcoin is the crypto which is the most stable according to Figure \ref{fig:daily1}.

\begin{figure}[h!]
    \centering
    \includegraphics[width=\linewidth]{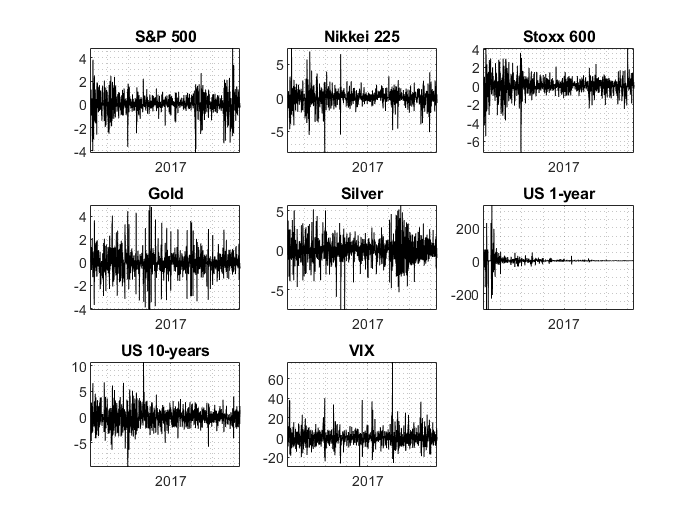}
    \captionof{figure}{Daily log returns of the eight crypto-predictors from August 8, 2015 till February 28, 2019.}
    \label{fig:xvar}
\end{figure}

The crypto market is 24/7 open, however the predictor variables are not. For this reason the data has to be adapted to use it for forecasting. The procedure is simple, when the market is closed for a variable the previous value of that variable is used. This gives a return of zero, however this is the best way since the variable is actually not changing for a day. Figure \ref{fig:xvar} shows the plots of the predictor variables.

\section{Methodology}
Studies have provided strong evidence of time-varying volatility in macroeconomic variables, however VARs with constant volatility are used in this paper. By using constant volatility the performance of point forecasting should not be affected that much by conditional heteroscedasticity, which is the case for heteroscedastic models like the GARCH and Stochastic Volatility. Heteroscedasticity is a major concern in the regression analysis, and in the analysis of variance, as it can invalidate statistical tests. These tests assume that the errors, obtained by modelling, are uniform and uncorrelated. For example the ordinary least squares (OLS) estimator is still unbiased in case of heteroscedasticity, it is inefficient because the actual variance and covariance are underestimated.  

In this paper, three types of specifications will be analysed the standard VAR model, the VAR with stochastic volatility and VAR with GARCH. The reason for multiple specifications of the model is to really see if the forecasting performance of a more complex model is better than a simple model. The Bayesian approach gives some advantages, the parameter uncertainty can be mitigated. The probabilistic statements can be computed without assumption. Another advantage is that the estimation of complex nonlinear models with many parameters is feasible. For the stochastic volatility there are two different models determined to investigate, one where the normal distribution is used and the other where the student-t distribution is used. These procedures by using these models are not the same so it could end up with different results. This way there can also be a conclusion about which distribution would give more accurate forecasts between all the models.

As stated in Catania, Grassi and Ravazzolo (2019), the number of lags of the VAR models is selected equal to three based on the BIC. The lag of interest of the cryptopredictors is the first lag. Thus, eight models will be discussed and used in this paper: Bayesian VAR(3), Bayesian VARX(3), Bayesian VAR(3)-SV, Bayesian VARX(3)-SV, Bayesian VAR(3)-GARCH, Bayesian VARX(3)-GARCH, Bayesian VAR(3)-SVt and Bayesian VARX(3)-SVt. These models are constant parameter vector autoregressive, these models are among the most common models applied in financial and macroeconomic forecasting, see Lutkepohl (2007) and Koop and Korobilis (2010). Regarding time-varying parameters, we left this issue as future research. To compare the models with each other, the Bayesian VAR(3) is chosen to be the benchmark. In the next subsections, the models used for the insample analysis and the forecasting exercise are explained briefly. 

\subsection{Bayesian VAR}
First of all, the focus is on the benchmark model, the Bayesian VAR(3) model is described as follows:
\begin{equation*}
    y_{t} = \beta_{1} y_{t-1} + \beta_{2} y_{t-2} + \beta_{3} y_{t-3} + \epsilon_{t}, \quad \epsilon_{t} \sim N(0,\Sigma_{\epsilon_{t}}), \text{ for } t=1,...,T,
\end{equation*}
\noindent with $T$ the number of total days of the data. Since this model is for every cryptocurrency the equation above can be rewritten in stacked form:
\begin{equation*}
    \begin{split}
        &Y_{t} = Z_{t} \beta + \epsilon_{t}, \quad \beta = vec({\beta_{1},\beta_{2},\beta_{3}}), \\
        & Z_{t} = (I_N \otimes X_{t}),
    \end{split}
\end{equation*}
\noindent where $X_t = [y_{t-1}, y_{t-2}, y_{t-3}]'$, for every cryptocurrency.

\subsubsection{Bayesian VARX}
In order to introduce possible dependence to other variables, it is possible to extend the Bayesian VAR model, by including other variables of interest. The so-called VARX model can be described as:

\begin{equation*}
    y_{t} = \beta_{1} y_{t-1} + \beta_{2} y_{t-2} + \beta_{3} y_{t-3} + \sum_{j=1}^8\gamma_{j} W_{j,t} + \epsilon_{i,t}, \quad \epsilon_{t} \sim N(0,\Sigma_{\epsilon_{t}}), \text{ for } t = 1,...,T,
\end{equation*}
\noindent with $T$ the number of total days of the data and where $\gamma_j$ and $W_{j,t}$ are the parameter and crypto-predictor respectively. Since this model is for every cryptocurrency the equation above can be rewritten in stacked form:
\begin{equation*}
    \begin{split}
        &Y_{t} = Z_t \beta + \epsilon_t, \quad \beta = vec({\beta_1,\beta_2,\beta_3,\gamma_1,...,\gamma_8}), \\
        & Z_{t} = (I_N \otimes X_{t}),
    \end{split}
\end{equation*}
\noindent with $T$ the number of total days of the data and where $X_t = [y_{t-1}, y_{t-2}, y_{t-3}, W_{1t},...,W_{8t}]'$, for every cryptocurrency.

\subsection{Bayesian VAR-SV}
In the following section, the models with time-varying volatility will be described in details by differenciating between SV and GARCH. First of all, the Bayesian VAR(3) with Stochastic Volatility is similar to the previous model, however there is a difference in the innovations term. This allows the model to take different approaches over time, for example in times of high uncertainty there could be a higher variance in the innovations. For this reason one should use stochastic volatility, since the model adapts to the movement and volatility of the time series. 

The Bayesian VAR-SV(3) model is described in the following way:
\begin{equation*}
    \begin{split}
        y_{t} &= \beta_{1} y_{t-1} + \beta_{2} y_{t-2} + \beta_{3} y_{t-3} + \epsilon_{t}, \\
        \epsilon_{t} &= A^{-1} \Lambda_{t}^{0.5}\varepsilon_{t}, \varepsilon_{t} \sim N(0,I_k), \Lambda_{t} \equiv \text{diag}(\lambda_{1t},...,\lambda_{kt}),\\
        \log(\lambda_{t}) &= \log(\lambda_{t-1}) + \nu_{t},\\
        \nu_{t} &= (\nu_{1t},\nu_{2t},...,\nu_{kt})' \sim N(0,\Phi), \text{ for } t = 1,...,T
    \end{split}
\end{equation*}
\noindent with $T$ the number of total days of the data and where $A$ is a lower triangular matrix with non-zero coefficients below the diagonal which are ones, $\Lambda_{t}$ is a diagonal matrix which contains the time-varying variances of shocks. This model implies that the reduced form variance-covariance matrix of innovations to the VAR is $var(\epsilon_{t}) \equiv \Sigma_{t} = A^{-1} \Lambda_{t} (A^{-1})'$ (Clark and Ravazzolo (2015)).


\subsection{Bayesian VAR-GARCH}
The Bayesian VAR(3) with GARCH(1,1) innovations is almost the same as the VAR-SV model, however there is a difference in the innovations term. This allows the model to take different approaches over time, for example in times of high uncertainty there could be a higher variance in the errors. It also has a memory over time so it can compare the observations with the past to get a better estimate of the predictions. For this reason one should use GARCH over SV, because of the memory over time. 

The Bayesian VAR(3) with GARCH(1,1) innovations is described in the following way:
\begin{equation*}
    \begin{split}
        y_{t} &= \beta_{1} y_{t-1} + \beta_{2} y_{t-2} + \beta_{3} y_{t-3} + \epsilon_{t}, \\
        \epsilon_{t} &= H_{t}^{0.5}\eta_{t}, \eta_{t} \sim N(0,I_k), H_{t} =  D_{t} R_{t} D_{t}, D_{t} = \text{diag}(h^{0.5}_{1t},...,h^{0.5}_{kt}),\\
        h_{t} &= \omega + B\epsilon^{(2)}_{t-1} + G h_{t-1}, \text{ for } t = 1,...,T,
    \end{split}
\end{equation*}
\noindent with $T$ the number of total days of the data and where $R$ is the conditional correlation matrix, $h_{t}$ follows a GARCH(1,1) model where $h_t = [h_{1t}, h_{2t}, ..., h_{kt}]'$ and $\epsilon_{t}^{(2)} = [\epsilon_{1t}^2, \epsilon_{2t}^2, ...,\epsilon_{kt}^2]'$ are conditional variances and squared errors respectively, $\omega$ and $B$ and $G$ are matrices of coefficients (Carnero and Eratalay (2014)).


\subsection{Bayesian VAR-SVt}
The following model description is similar to the VAR-SV, but now with a student-t distribution. This model is in this paper referred to as VAR-SVt and described as:
\begin{equation*}
    \begin{split}
        y_{t} &= \beta_{1} y_{t-1} + \beta_{2} y_{t-2} + \beta_{3} y_{t-3} + \epsilon_{t}, \\
        \epsilon_{t} &= A^{-1} \Lambda_{t}^{0.5}\varepsilon_{t}, \varepsilon_{t} \sim t(0,I_k,\eta), \Lambda_{t} \equiv \text{diag}(\lambda_{1t},...,\lambda_{kt}),\\
        \log(\lambda_{t}) &= \log(\lambda_{t-1}) + \nu_{t},\\
        \nu_{t} &= (\nu_{1t},\nu_{2t},...,\nu_{kt})' \sim t(0,\Phi,\eta), \text{ for } t = 1,...,T
    \end{split}
\end{equation*}
\noindent with $T$ the number of total days of the data, and $\eta$ the degrees of freedom. $A$ is a lower triangular matrix with non-zero coefficients below the diagonal which are ones, $\Lambda_{t}$ is a diagonal matrix which contains the time-varying variances of shocks. This model implies that the reduced form variance-covariance matrix of innovations to the VAR is $var(\epsilon_{t}) \equiv \Sigma_{t} = A^{-1} \Lambda_{t} (A^{-1})'$ (Clark and Ravazzolo (2015)).


\subsection{Forecasting}
To forecast the cryptocurrencies the methodology used is called a rolling window. The estimation part will be from 08/08/2015 till 08/08/2017 so a two year estimation window. Using the results from this estimation the point forecast one-day ahead will be calculated. The next forecast will be done by estimation a day later than before, so from 09/08/2015 till 09/08/2017. This procedure will proceed until the end of the data is reached (02/28/2019) which is after 567 days, thus the number of one-day ahead forecasts is 567. As a prior for the SV and GARCH models the Minnesota prior is used as a start. This approach is standard and can be extended to other priors, for this paper the standard approach is sufficient enough to investigate the cryptocurrencies. For every one-day forecast, a total of 6000 simulations are drawn and the first 1000 simulations are burned. This burning of the first simulation is due to the fact that the first simulations can be correlated and/or inaccurate. Over time the simulations are independent of each other and can be used for measures. 

\subsection{Measures}
To compare the performances of the forecasts, we will use five different types of measures. The first three are measures of point forecasts, the last two are measures of density forecasts. The difference between measures using point forecasts and measures using density forecasts is that measures using point forecasts uses the mean of the simulations, however for measures using density forecasts all the simulations are used. Measures using density forecasts will give a great view of the full simulation and will not be averaged out like the measures using point forecasts.  However measures using point forecasts still give a good interpretation of the performance and is more efficient in time. 

First measure is the so-called 95\% credible interval, this is an interval obtained by simulations. The 2.5\% and 97.5\% quantile's of the simulations are the lower and upper bound respectively. The idea behind this credible interval is that in 95\% of the cases the forecast will be in this interval. Another measure is the sign predictability, in this paper referred as the 'Success rate', the percentage of the forecasts which are in the right direction, as the actual observations. When the actual observation goes down and the forecast as well then it counts as a 'success', it is also a 'success' when the actual observation goes up and the forecast as well. In the two other cases it counts as a 'fail', in this way the 'Success rate' is built.  We do not perform sign predictability tests, the reason for this was indicated by Christoffersen and Diebold (2006). Tests that rely on the sign gives no information about volatility dynamics, which is potentially valuable for detecting sign predictability. 

The third measure is called the Root Mean Squared Error (RMSE). The RMSE is preferred over the Mean Squared Error (MSE) since it is on the same scale as the data. Some authors (e.g., Armstrong, 2001) recommend the use of the RMSE since it is more sensitive to outliers than commonly used Mean Absolute Error (MAE). The RMSE is computed for each cryptocurrency series, $i$ = Bitcoin, Ethereum, Ripple and Litecoin:

\begin{equation*}
    \text{RMSE}_i = \sqrt{\frac{\sum^{T-1}_{t=R}({\hat{y}_{i,t+1}}-y_{i,t+1})^2}{T-R}}
\end{equation*}

\noindent where $R$ is the length of the rolling window, $T$ the number of observations, $\hat{y}_{i,t+1}$ the $i$th-cryptocurrency forecast at time $t$, and $y_{i,t+1}$ is the actual observation at time $t$. 

The fourth type of measure is for evaluating the density forecasts, this measure is called the Log Predictive Score (LS). In the same way as for the RMSE, it is computed for each series:

\begin{equation*}
    \text{LS}_i = \sum^{T-1}_{t=R}\ln{f(y_{i,t+1})}
\end{equation*}

\noindent where $f(y_{i,t+1})$ is the predictive density for $y_{i,t+1}$, given the information up to time $t$. The fifth measure is the Continuous Rank Probability Score (CRPS). This is a continuous extension of the RPS and can be defined by considering an integral of the Brier scores over all possible thresholds $x$. Denoting the predicted cumulative density function by $F(x) = p(X \leq x)$ and the observed value of $X$ by $y_i$, the continuous ranked probability score can be written for each series as:

\begin{equation*}
    \text{CRPS}_i = E \left( \int^{\infty}_{-\infty} [F(x) - H(x-y_i)]^2 dx \right),
\end{equation*}
\noindent where $H(x-y_i)$ is the Heaviside function that takes the value 0 when the observed value is smaller than the threshold, and 1 otherwise (Jolliffe and Stephenson, 2003, Forecast Verification). 

For the RMSE, LS and CRPS we apply a $t$-test by Diebold and Mariano (1995) of each model versus the benchmark. This test gives a p-value which indicates a certain significance level. If in a table a value has one star then the model performs better, by a significance level of 5\%, than the benchmark model. If in a table a value has two stars then the model performs better, by a significance level of 1\%, than the benchmark model. The first row of the tables contain the results of the RMSE, LS and CRPS of the benchmark which is the BVAR model. Ratios of each models RMSE and CRPS to the benchmark are done such that entries less than 1 indicate that the given model yields forecasts more accurate than those from the benchmark. While for the differences of each models LS to the benchmark are done such that a positive number indicates a model beats the baseline.

The other procedure we use is the model confidence set procedure of Hansen, Lunde, and Nason (2011) using a $R$ package called:$MCS$, detailed by Bernardi and Catania (2016). The model confidence set procedure compares all the predictions jointly and deletes a model if it is significantly worse, one end up with the best possible models of the models that were put in. The models which have a grey background in tables will be chosen to be not significantly worse than the other models.

\section{Results}
As stated in section 4.6, we use different measures for point and density forecasting. Initially the focus will be set on point forecasting. The first results of the forecasts are given in Table \ref{table:95bti}, these are the percentages of actual observations  outside of the 95\% credible interval obtained by simulation. To compare the BVAR model with the BVAR-GARCH model the forecasts of the BVAR-GARCH model is only for the Ripple not more often in the 95\% credible interval. This would imply that the forecasts are less volatile using the BVAR-GARCH model compared to the BVAR model, for the Ripple this would be the opposite. This is in line with the expectations since the kurtosis of the Ripple (see Table \ref{table:statistics}) is significantly higher than the other cryptocurrencies. The BVAR-SV and BVARX-SV models have the highest percentages of all the cryptocurrencies except for the Bitcoin. This would suggest that using Stochastic Volatility will not give a good prediction overall using credible intervals.
The results between the BVAR model and the BVARX model are close to each other, so there is not a clear distinction between these two models. However the BVARX-GARCH model is the model that stands out the most, this gives the most forecasts in the 95\% credible interval, the only exceptions are the BVAR-GARCH model for Ethereum and the BVARX-SV model for Bitcoin. 

Overall the use of the crypto-predictor variables would be helpful to simulate forecasts due to the fact that in almost every case using the crypto-predictor variables would give a lower percentage of actual observations outside of the 95\% credible interval. Using a student-t distribution in the SV model is only for Bitcoin more often out of the interval, which is expected as Bitcoin is the least volatile of the cryptocurrencies. Including the crypto-predictor variables into the SV-t model this percentage is only smaller for the Ripple, however not by a lot.

\begin{table}[h!]
\centering
\caption{Percentage of actual observations outside of the 95\% credible interval retrieved by simulation}
\begin{tabular}{l l l l l}
\hline
Cryptocurrency & Bitcoin & Ethereum & Ripple & Litecoin \\ \hline
BVAR & 8.9947 & 5.1146 & 4.7619 & 6.5256 \\
BVAR-SV & 5.8201 & 21.517 & 14.991 & 16.755 \\
BVAR-GARCH & 5.9965 & 3.7037 & 5.4674 & 4.4092 \\
BVARX & 9.1711 & 4.5855 & 4.9383 & 6.7019 \\
BVARX-SV & 3.5273 & 13.404 & 8.9947 & 8.9947 \\
BVARX-GARCH & 5.6437 & 4.0564 & 4.0564 & 3.351\\
BVAR-SVt & 7.7601 & 6.5256 & 9.7002 & 10.582 \\
BVARX-SVt & 8.1129 & 6.3492 & 9.1711 & 10.582 \\
\hline
\label{table:95bti}
\end{tabular}
\end{table}

For every cryptocurrency also the credible intervals are plotted (see Figures \ref{fig:95bit}-\ref{fig:95rip} in the Appendix). In these figures the credible interval of the BVAR models are pretty steady for all cryptocurrencies, hence that these models are not capturing the volatile movements of the data that well. When one uses a more expanded version like the BVAR-SV or BVAR-GARCH model the credible levels captures the movements better, when there are shocks the credible levels adapt to its movement. However the BVARX-SV models stands out the most, there is much noise in the credible levels, so using the predictors would not be helpful to give a more narrow credible interval to predict one day ahead. 

Table \ref{table:rightdir} shows the results for the second point forecasting measure previously described. This predictability is not statistically tested but gives an insight into the accuracy of the movement of the forecasts. The returns are used to see if the direction of predictions is correct. The BVAR-SV model is compared to the BVAR model and BVAR-GARCH model in all cases more in the right direction. Another observation is that only for Ethereum and the Ripple including the crypto-predictor variables predict the direction more precise. The reason for this behaviour would be that Ripple is more dependent on market movement than the other cryptocurrencies. However the percentages are under 50\% or close to 50\% which would imply that these models (BVAR and BVAR-GARCH) cannot predict the movement very precise. That statement only applies for now on the prediction of the cryptocurrency going up or down. 

An important observation of this table is that the stochastic volatility models have the best scores overall and are in some cases about 60-67\% which is much more precise than for example 35.45\% of the BVAR-GARCH for the Bitcoin. This is especially the case for the SV model with a student-t distribution, thus using a SV model with student-t distribution is the best way, among these models, to forecast the direction of the cryptocurrencies.

\begin{table}[h!]
\centering
\caption{Percentage of forecasts in the right direction (up or down)}
\begin{tabular}{l l l l l}
\hline
Cryptocurrency & Bitcoin & Ethereum & Ripple & Litecoin \\ \hline
BVAR & 51.675 & 43.563 & 48.325 & 44.621 \\
BVAR-SV & 51.852 & 55.556 & 55.556 & 55.732 \\
BVAR-GARCH & 35.45 & 37.39 & 38.801 & 38.448 \\
BVARX & 47.795 & 45.15 & 49.735 & 43.034 \\
BVARX-SV & 51.852 & 56.085 & 56.614 & 50.794 \\
BVARX-GARCH & 35.097 & 41.446 & 41.975 & 36.861 \\
BVAR-SVt & 61.905 & 62.963 & 61.905 & 67.901 \\ 
BVARX-SVt & 62.434 & 62.963 & 58.025 & 67.725 \\ 
\hline
\label{table:rightdir}
\end{tabular}
\end{table}

Moving to the last point forecast measure Table \ref{table:ratiormse} contains the results of the ratio of the RMSE. For these results, the RMSE of the benchmark model (BVAR) and the ratios of the other models are reported. As expected in the descriptive statistics the Ripple is the cryptocurrency with the highest RMSE due to the high kurtosis. 

For Ripple and Litecoin the SV models are significantly better than the benchmark model. The GARCH model is in all cases not significantly better than the benchmark, the cause could be that cryptocurrencies do not follow such dynamics. We could state that including the crypto-predictor variables is not affecting the RMSE of the models enough to increase the performance of the forecasts. For the Bitcoin there is no model significantly better performing than the VAR, this could be caused by the aforementioned stability of the Bitcoin compared to the other cryptocurrencies.

\begin{table}[h!]
\centering
\caption{Ratio of RMSE against benchmark}
\begin{threeparttable}
\centering
\begin{tabular}{l l l l l}
\hline
Cryptocurrency & Bitcoin & Ethereum & Ripple & Litecoin \\ \hline
BVAR & \cellcolor[gray]{0.8}4.6091 & \cellcolor[gray]{0.8}5.6996 & 7.6627 & 6.7055 \\ \hline
BVAR-SV vs BVAR & \cellcolor[gray]{0.8}0.99466 & \cellcolor[gray]{0.8}0.99466 & \cellcolor[gray]{0.8}0.98465** & \cellcolor[gray]{0.8}0.97735** \\
BVAR-GARCH vs BVAR & 1.0072 & 1.0106 & 1.0189 & 1.0163 \\
BVARX vs BVAR & 1.0111 & 1.0113 & 1.0057 & 1.0098 \\
BVARX-SV vs BVAR & \cellcolor[gray]{0.8}0.99585 & \cellcolor[gray]{0.8}0.99598 & \cellcolor[gray]{0.8}0.98555** & 0.98187* \\
BVARX-GARCH vs BVAR & 1.013 & 1.02 & 0.99486 & 1.0065 \\
BVAR-SVt vs BVAR & \cellcolor[gray]{0.8}0.99593 & \cellcolor[gray]{0.8}0.98915** & \cellcolor[gray]{0.8}0.98254** & \cellcolor[gray]{0.8}0.98709** \\
BVARX-SVt vs BVAR & \cellcolor[gray]{0.8}0.99744 & \cellcolor[gray]{0.8}0.98927* & \cellcolor[gray]{0.8}0.98349** & \cellcolor[gray]{0.8}0.98774** \\
\hline
\end{tabular}
\vspace{-.2cm}
\textit{Notes:}
\begin{tablenotes} 
\item[1] The 'X' indicates models with the crypto-predictor variables included, the 't' indicates that the student-t distribution is used.
\item[2] For BVAR, the benchmark model, the table reports the RMSE, for other models it reports the ratio between the RMSE of the current model and the benchmark. Entries less than 1 indicate that forecasts from current model are more accurate than forecasts from the benchmark model.
\item[3] ** and * indicate RMSE ratios are significantly different from 1 at 5\% and 10\%, according to the Diebold-Mariano test.
\item[4] Gray cells indicate models that belong to the Superior Set of Models delivered by the Model Confidence Set procedure at confidence level 10\%.
\end{tablenotes}
\end{threeparttable}
\label{table:ratiormse}
\end{table}

The grey areas indicate the model confidence set, this also confirms our conclusion that using the SV model is in almost every case (except for Litecoin the VARX-SV) in this set. If one wants to forecast these cryptocurrencies with one of these models, then the preferred option, by looking at the RMSE, is using stochastic volatility. 

The last two tables (see Tables \ref{table:ratiocrps} and \ref{table:ratiopl}) contain the results of the density measures CRPS and PL. The results of the CRPS measure are not that different from the RMSE. One difference is that by the CRPS, GARCH outperforms the VAR for the Bitcoin and for the Ripple if the crypto-predictor variables are included. Hence the density of the Bitcoin and Ripple follow the dynamics of a GARCH model more than the benchmark. However the SV model also outperforms the GARCH model since the values of the SV model are in many cases lower. In the model confidence set is now also the GARCH for the Bitcoin included.

\begin{table}[h!]
\centering
\caption{Ratio of CRPS against benchmark}
\begin{threeparttable}
\begin{tabular}{l l l l l}
\hline
Cryptocurrency & Bitcoin & Ethereum & Ripple & Litecoin \\ \hline
BVAR & 2.4707 & \cellcolor[gray]{0.8}3.1043 & 3.9479 & \cellcolor[gray]{0.8}3.453 \\ \hline
BVAR-SV vs BVAR & \cellcolor[gray]{0.8}0.95108** & \cellcolor[gray]{0.8}0.99346 & 0.90827** & \cellcolor[gray]{0.8}0.9735* \\
BVAR-GARCH vs BVAR & \cellcolor[gray]{0.8}0.96574** & 1.0443 & 0.99732 & 1.0226 \\
BVARX vs BVAR & 1.0125 & 1.012 & 1.007 & 1.0131 \\
BVARX-SV vs BVAR & 1.066 & 1.0298 & 0.93993** & \cellcolor[gray]{0.8}0.99681 \\
BVARX-GARCH vs BVAR & 0.97812* & 1.042 & 0.97615* & 1.0216 \\
BVAR-SVt vs BVAR & \cellcolor[gray]{0.8}0.95964** & \cellcolor[gray]{0.8}0.98594 & \cellcolor[gray]{0.8}0.88674** & \cellcolor[gray]{0.8}0.96403** \\
BVARX-SVt vs BVAR & \cellcolor[gray]{0.8}0.96002** & \cellcolor[gray]{0.8}0.98764 & \cellcolor[gray]{0.8}0.88773** & \cellcolor[gray]{0.8}0.96525** \\
\hline
\end{tabular}
\vspace{-.2cm}
\textit{Notes:}
\begin{tablenotes} 
\item[1] The 'X' indicates models with the crypto-predictor variables included, the 't' indicates that the student-t distribution is used.
\item[2] For BVAR, the benchmark model, the table reports the CRPS, for other models it reports the ratio between the CRPS of the current model and the benchmark. Entries less than 1 indicate that forecasts from current model are more accurate than forecasts from the benchmark model.
\item[3] ** and * indicate CRPS ratios are significantly different from 1 at 5\% and 10\%, according to the Diebold-Mariano test.
\item[4] Gray cells indicate models that belong to the Superior Set of Models delivered by the Model Confidence Set procedure at confidence level 10\%.
\end{tablenotes}
\end{threeparttable}
\label{table:ratiocrps}
\end{table}

The conclusion drawn from the first measure of density forecast (CRPS) is that for Ethereum the case is now the same as the case for Bitcoin by using the RMSE, there is no model significantly better than the benchmark. The reason could be that the density of the forecasts of  Ethereum are not following the movement captured by the used models, such that the predictability of Ethereum is low caused by uncertainty higher than the other cryptocurrencies. 

Regarding the density forecast for CRPS, the main conclusion is that including stochastic volatility in the model formulation lead to better results with respect to the benchmark (VAR model) and to GARCH specification. In particular the inclusion of student-t specification of the errors in the SV models leads to better results and to great improvements for every cryptocurrency. If one includes the crypto-predictors in the analysis, there are not so great improvements except when the errors are student-t specified for stochastic volatility.

\begin{table}[h!]
\centering
\caption{Differences of PL against benchmark}
\begin{threeparttable}
\begin{tabular}{l l l l l}
\hline
Cryptocurrency & Bitcoin & Ethereum & Ripple & Litecoin \\ \hline
BVAR & \cellcolor[gray]{0.8}-3.2676 & -3.1777 & \cellcolor[gray]{0.8}-3.7552 & \cellcolor[gray]{0.8}-3.8476 \\ \hline
BVAR-SV vs BVAR & \cellcolor[gray]{0.8}0.28254 & \cellcolor[gray]{0.8}-1.6413** & \cellcolor[gray]{0.8}-0.030439 & \cellcolor[gray]{0.8}-0.17147 \\
BVAR-GARCH vs BVAR & \cellcolor[gray]{0.8}-0.081207 & \cellcolor[gray]{0.8}-0.76657* & \cellcolor[gray]{0.8}0.27199 & \cellcolor[gray]{0.8}0.27338 \\
BVARX vs BVAR & \cellcolor[gray]{0.8}-0.023045 & -0.0085887* & \cellcolor[gray]{0.8}-0.025829 & \cellcolor[gray]{0.8}-0.027074 \\
BVARX-SV vs BVAR &  \cellcolor[gray]{0.8}0.28375 & \cellcolor[gray]{0.8}-0.85084** & \cellcolor[gray]{0.8}0.25762 & \cellcolor[gray]{0.8}0.39481 \\
BVARX-GARCH vs BVAR & \cellcolor[gray]{0.8}0.239 & -0.27849 & 0.3654 & \cellcolor[gray]{0.8}0.40684 \\
BVAR-SVt vs BVAR & 0.38974 & 0.067936  & 0.59546 & 0.63765 \\
BVARX-SVt vs BVAR & 0.43121 & 0.064834 & 0.55927 & \cellcolor[gray]{0.8}0.4663 \\
\hline
\end{tabular}
\vspace{-.2cm}
\textit{Notes:}
\begin{tablenotes} 
\item[1] The 'X' indicates models with the crypto-predictor variables included, the 't' indicates that the student-t distribution is used.
\item[2] For BVAR, the benchmark model, the table reports the PL, for other models it reports the difference between the PL of the current model and the benchmark. Entries greater than 0 indicate that forecasts from current model are more accurate than forecasts from the benchmark model.
\item[3] ** and * indicate PL differences are significantly different from 0 at 5\% and 10\%, according to the Diebold-Mariano test.
\item[4] Gray cells indicate models that belong to the Superior Set of Models delivered by the Model Confidence Set procedure at confidence level 10\%.
\end{tablenotes}
\end{threeparttable}
\label{table:ratiopl}
\end{table}

The predictive likelihood (PL, or log predictive score (LS)) has some different results compared to the previous measures. At first the predictive likelihood is very close to each other if one compares the cryptocurrencies, this indicates that the models perform the same for the cryptocurrencies. Only for Ethereum there are models significantly better performing than the VAR. The SV models are in that case the most significant and the GARCH and the VAR including the crypto-predictor variables are less significant. 

Overall the model confidence set is as before containing the SV models. However this time the SV-t models are not in this set, only for the Litecoin including the crypto-predictor variables. Litecoin has however almost a full set, only the SV-t model is not in it, so Litecoin is not following a single model, but can be explained by multiple. The GARCH models are now in the model confidence set as well, this illustrates that the log score of the forecasts are describable as GARCH movements.

Regarding the density forecast for PL, the main conclusion is that including stochastic volatility in the model formulation lead for Ethereum to better results with respect to the benchmark (VAR model) and to GARCH specification. In contrary to the CRPS inclusion of the student-t specification of the errors in the SV model lead to no significant better results. If one includes crypto-predictors in the analysis, there are only for Ethereum improvements if there is no student-t specification.

\subsection{Robustness check}

In this section, we perform the forecasting exercises by including different univariate models. We report the results for different possible benchmark models. We consider the following two univariate models: an autoregressive model with one lag (AR(1)) and an autoregressive model with the first three lags (AR(3)) based on the BIC criterion. 

Table \ref{Tab_Uni} reports the point and density forecasting for the AR(1) and AR(3) versus the benchmark model considered in Section 5. All these models are run by using the usual Bayesian priors and by running $5.000$ iterations. Furthermore, we perform the root mean square error (RMSE) and the CRPS for the four main cryptocurrencies. As stated in Table \ref{Tab_Uni} the results for the point and density forecasting are qualitatively similar to multivariate benchmark case, VAR(3). 

\begin{table}[h!]
\centering
\begin{tabular}{lcccc}
\hline
Models & Bitcoin & Ethereum & Ripple & Litecoin \\
\hline
\textit{RMSE} \\
BAR(1) & 4.6033  &  5.6470  &  7.5795 &   6.5794 \\
BAR(3) & 4.6069  &  5.6517  &  7.5984  &  6.6076 \\
BVAR(3) & 4.6091 & 5.6996 & 7.6627 & 6.7055 \\
\hline
\textit{CRPS} \\
BAR(1) & 2.4717 &   3.0790 &   3.8395 &   3.4161  \\
BAR(3) &  2.4730  &  3.0809  &  3.8816  &  3.4245 \\
BVAR(3) &  2.4707 & 3.1043 & 3.9479 & 3.453 \\
\hline
\end{tabular}
\caption{Point (RMSE) and Density forecasting (CRPS) for Bayesian AR(1), AR(3) and VAR(3).}
\label{Tab_Uni}
\end{table}

\section{Conclusion}\label{s:conclusion}
Recently cryptocurrencies have attracted attention from researchers and financial institutions due to their importance. In this paper a comparison of the performance of several models has been investigated to predict four of the most capitalised cryptocurrencies: Bitcoin, Ethereum, Ripple and Litecoin. A set of crypto-predictors is applied and eight model combinations are proposed for combining these predictors. The results show statistically significant improvements in point forecasting for all the cryptocurrencies when using a combination of stochastic volatility and a student-t distribution. In density forecasting for all cryptocurrencies the stochastic volatility model gives the best predictability. 
One recommendation for future research is to allow different weights across time and time-varying parameters to improve the point and density forecasting. Moreover, other crypto-predictors based on the dynamics of the crypto-market might be interesting for modeling.

\section*{Acknowledgments} 

Luca Rossini acknowledges financial support from the European Union Horizon 2020 research and innovation programme under the Marie Sklodowska--Curie grant agreement No 796902. 

\clearpage

\renewcommand{\thesection}{A}
\renewcommand{\theequation}{A.\arabic{equation}}
\renewcommand{\thefigure}{A.\arabic{figure}}
\renewcommand{\thetable}{A.\arabic{table}}
\setcounter{table}{0}
\setcounter{figure}{0}
\setcounter{equation}{0}

\section{Supplementary Material} \label{s:appendix}
\subsection{Results}

\begin{figure}[h!]
  \centering
  \includegraphics[width=\linewidth]{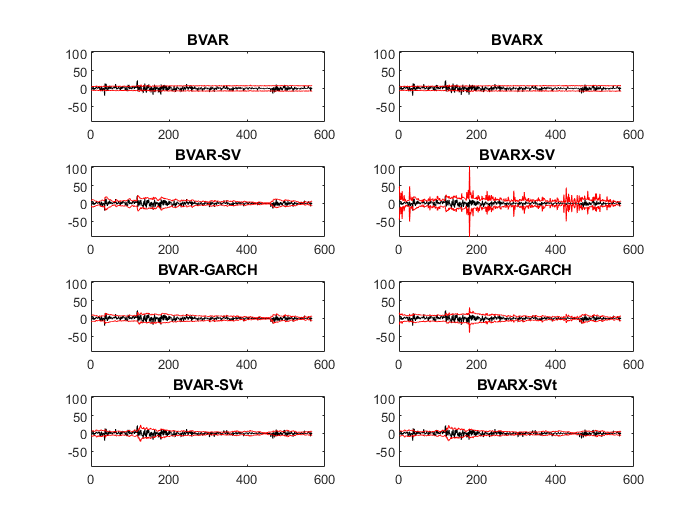}
  \caption{Credible interval for Bitcoin}
  \label{fig:95bit}
\end{figure}

\begin{figure}[h!]
  \centering
  \includegraphics[width=\linewidth]{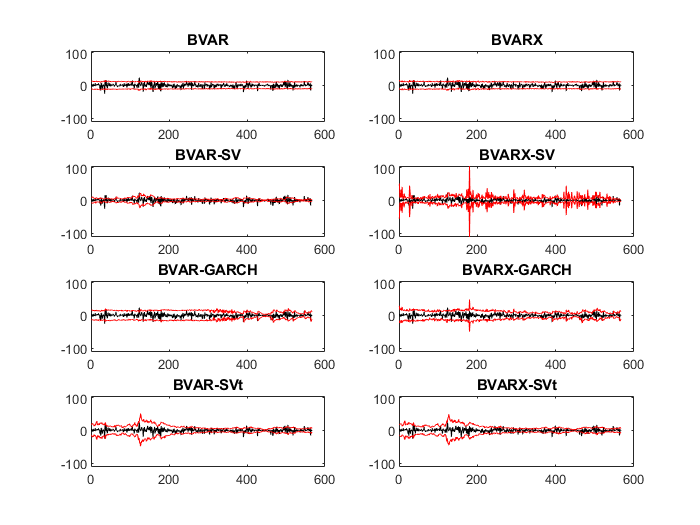}
  \caption{Credible interval for Ethereum}
  \label{fig:95eth}
\end{figure}

\begin{figure}[h!]
  \centering
  \includegraphics[width=\linewidth]{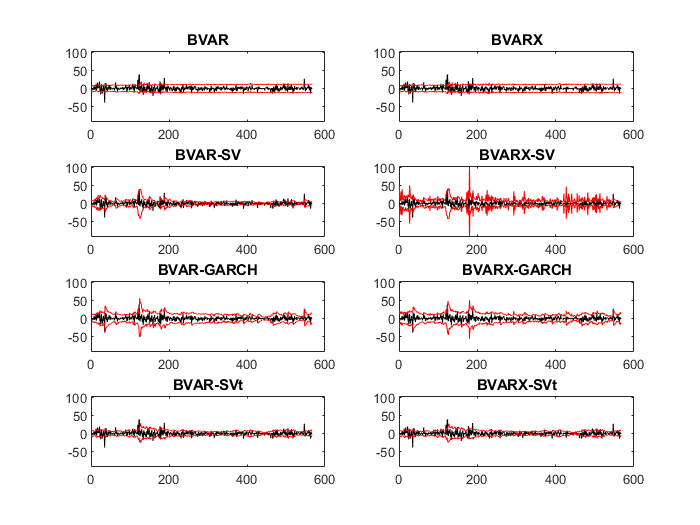}
  \caption{Credible interval for Litecoin}
  \label{fig:95lit}
\end{figure}

\begin{figure}[h!]
  \centering
  \includegraphics[width=\linewidth]{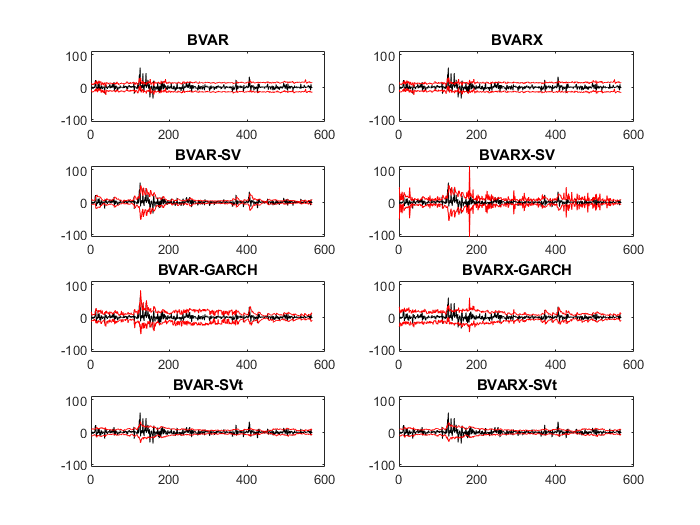}
  \caption{Credible interval for Ripple}
  \label{fig:95rip}
\end{figure}


\begin{thebibliography}{100}

\bibitem[Armstrong (2001)]{armstrong}
Armstrong J.S. (2001).
\textit{Evaluating Forecasting Methods}. In Principles of Forecasting: A Handbook for Researchers and Practitioners (Ed. J. Scott Armstrong). Kluwer.

\bibitem[Bernardi and Catania (2016)]{bernardi}
Bernardi M. and Catania L. (2016).
\textit{Portfolio optimisation under flexible dynamic dependence modelling}. ArXiv e-prints.

\bibitem[Bianchi (2018)]{bianchi} 
Bianchi D. (2018). 
\textit{Cryptocurrencies as an asset class? An empirical assessment}.
Tech. rep., SSRN working paper.

\bibitem[Bloomberg (2017a)]{bloomberg}
Bloomberg (2017a).
\textit{Japan’s BITpoint to add bitcoin payments to retail outlets}.\\ \href{https://www.bloomberg.com/news/articles/2017-05-29/japans-bitpoint-to-add-bitcoin-payments-to-100-000s-of-outlets.}{https://www.bloomberg.com/news/articles/2017-05-29/japans-bitpoint-to-add-bitcoin-payments-to-100-000s-of-outlets.}

\bibitem[Bloomberg (2017b)]{bloomberg2}
Bloomberg (2017b).
\textit{Some central banks are exploring the use of cryptocurrencies}.  \\
 \href{https://www.bloomberg.com/news/articles/201706-28/rise-of-digital-coins-has-central-banks-considering-eversions.}{https://www.bloomberg.com/news/articles/201706-28/rise-of-digital-coins-has-central-banks-considering-eversions.}

\bibitem[Bloomberg (2018)]{bloomberg3}
Bloomberg (2018).
\textit{Big Investors May Be Dragging Bitcoin Toward Market Correlation}.

\bibitem[Capgemini (2017)]{capgemini}
Capgemini, BNP Paribas. (2017).
\textit{World Payments Report 2017}. Payments volume. Retrieved from https://www.worldpaymentsreport.com/\#non-cash payments-content.

\bibitem[Carnero and Eratalay (2014)]{carnero}
Carnero M.A. and Eratalay M.H., (2014).
\textit{Estimating VAR-MGARCH models in multiple steps}, Studies in Nonlinear Dynamics and Econometrics, De Gruyter, vol. 18(3), pages 1-27.

\bibitem[Catania, Grassi and Ravazzolo (2018)]{catania}
Catania L. , Grassi S. , and Ravazzolo F. (2018).
In M. Corazza, M. Durbán, A. Grané, C. Perna, and M. Sibillo (Eds.),
\textit{Predicting the volatility of cryptocurrency time-series. In Mathematical and statistical methods for actuarial sciences and finance}, Charm: Springer.

\bibitem[Catania, Grassi and Ravazzolo (2019)]{catania2}
Catania L. , Grassi S. , and Ravazzolo F. (2019).
\textit{Forecasting cryptocurrencies under model and parameter instability}. International Joural of Forecasting, 35, 485-501.

\bibitem[Chan (2019)]{chan}
Chan J.C.C. (2019).
\textit{Large Bayesian Vector Autoregressions}.

\bibitem[Cheah and Fry (2015)]{cheah}
Cheah E. and Fry J. (2015).
\textit{Speculative bubbles in Bitcoin markets? An empirical investigation into the
fundamental value of Bitcoin}.
Economics Letters 130, 32–37.

\bibitem[Christoffersen and Diebold (2006)]{christ}
Christoffersen P. and Diebold F. (2006). 
\textit{Financial asset returns, direction-of-change forecasting, and volatility dynamics}. Management Science, 52, 1273–1287.

\bibitem[Chu, Nadarajah and Chan (2015)]{chu}
Chu J., Nadarajah S.  and Chan S. (2015).
\textit{Statistical analysis of the exchange rate of bitcoin}. PloS One, 10, 1–27.

\bibitem[Clark and Ravazzolo (2015)]{clark}
Clark T. E. and Ravazzolo F. (2015). 
\textit{Macroeconomic Forecasting Performance under Alternative Specifications of Time‐Varying Volatility}. Journal of Applied Econometrics Volume 30, Issue 4. 

\bibitem[Cointelegraph (2017)]{coin}
Cointelegraph (2017).
\textit{South Korea officially legalizes bitcoin, huge market for traders}.
https://cointelegraph.com/news/south-korea-officiallylegalizes-bitcoin-huge-market-for-traders.

\bibitem[D’Agostino, Gambetti and Giannone (2013)]{dagostino}
D’Agostino A., Gambetti L., Giannone D. (2013). 
\textit{Macroeconomic forecasting and structural change}.
Journal of Applied Econometrics 28, 82–101.

\bibitem[Diebold and Mariano (1995)]{diebold}
Diebold F. and Mariano R. (1995).
\textit{Comparing predictive accuracy}.
Journal of Business and Economic Stastistics, 13, 253–263.

\bibitem[Dyhrberg (2016)]{dyhrberg}
Dyhrberg A.H. (2016).
\textit{Hedging capabilities of bitcoin. Is it the virtual Gold?}
Finance Research Letters, 16, 139-144. 

\bibitem[Ethereum (2014)]{eth}
Ethereum (2014). Ethereum Wiki. https://github.com/ethereum/wiki/ wiki/White-Paper/.

\bibitem[Fernández-Villaverde and Sanches (2016)]{fernandez}
Fernández-Villaverde J. and Sanches D. (2016). 
\textit{Can Currency Competition Work?}
NBER Working Papers 22157. New York: National Bureau of Economic Research.

\bibitem[Forbes (2017)]{forbes}
Forbes (2017).
\textit{Emerging applications for blockchain}.\\
https://www.forbes.com/sites/forbestechcouncil/2017/07/18/emerging-applicationsfor-blockchain.

\bibitem[Griffin and Shams (2018)]{griffin}
Griffin J.M. and Shams A. (2018).
\textit{Is Bitcoin Really Un-Tethered?}
Technical Report, SSRN Working Paper. Rochester: SSRN.

\bibitem[Hansen, Lunde and Nason (2011)]{hansen}
Hansen P.R., Lunde A. , and Nason J.M. (2011).
\textit{The model confidence set}.
Econometrica, 79, 453–497.

\bibitem[Hencic and Gourieroux (2014)]{hencic}
Hencic A. and Gourieroux C. (2014).
\textit{Noncausal autoregressive model in application to bitcoin/USD exchange rate}.
In Proceedings of the 7th financial risks international forum (pp. 1–25).

\bibitem[Hotz-Behofsits, Huber and Zorner (2018)]{hotz}
Hotz-Behofsits C., Huber F. , and Zorner T.O. (2018).
\textit{Predicting cryptocurrencies using sparse non–gaussian state space models}. Journal of Forecast, 36(6), 627–640.

\bibitem[Johannes, Korteweg and Polson (2014)]{johannes}
Johannes M., Korteweg A. and Polson N. (2014).
\textit{Sequential learning, predictive regressions, and optimal portfolio returns}.
Journal of Finance, 69, 611–644.

\bibitem[Jolliffe and Stephenson (2003)]{jolliffe}
Jolliffe I.T. and Stephenson D.B. (2003).
\textit{Forecast Verification}. A Practitioner’s Guide in Atmospheric Science. John Wiley and Sons Ltd., Hoboken, 240 p. 

\bibitem[Koop and Korobilis (2010)]{koop}
Koop G. and Korobilis D. (2010).
\textit{Forecasting inflation using dynamic model averaging}. Bayesian Multivariate Time Series Methods for Empirical Macroeconomics, 3, 267–358.

\bibitem[Koop and Korobilis (2013)]{koop2}
Koop G. and Korobilis D. (2013).
\textit{Large time-varying parameter VARs}.
Journal of Econometrics, 177, 185–198.

\bibitem[Litecoin (2014)]{lit}
Litecoin (2014). Litecoin Wiki. https://litecoin.info/Litecoin/. 

\bibitem[Lutkepohl (2007)]{lutkepohl}
Lutkepohl H. (2007).
\textit{New introduction to multiple time series analysis}.
Springer Publishing Company, Incorporated.

\bibitem[Muglia, Santabarbara and Grassi (2019)]{muglia}
Muglia C., Santabarbara L. and Grassi S. (2019).
\textit{Is Bitcoin a Relevant Predictor of Standard \& Poor’s 500?}
J. Risk Financial Manag. 2019, 12(2), 93.

\bibitem[Nakomoto (2008)]{nakomoto}
Nakomoto S. (2008).
\textit{Bitcoin: A Peer-to-Peer Electronic Cash System}.

\bibitem[Raftery, Kárn\`y and Ettler (2010)]{raftery}
Raftery A. E., Kárn\`y M. , and Ettler P. (2010).
\textit{Online prediction under model uncertainty via dynamic model averaging: Application to a cold  rolling mill}.
Technometrics, 52, 52–66.

\bibitem[Ripple (2012)]{rip}
Ripple (2012). Welcome to Ripple. https://ripple.com/. 

\bibitem[Sapuric and Kokkinaki (2014)]{sapuric}
Sapuric S. and Kokkinaki A. (2014). \textit{Bitcoin is volatile! Isn’t that right? Business Information Systems Workshops}, 255–265.

\bibitem[Sims (1980)]{sims}
Sims C.A. (1980).
\textit{Macroeconomics and Reality}. 
Econometrica, Vol. 48, No. 1. (Jan., 1980), pp. 1-48.

\bibitem[Sims and Zha (2006)]{sims2}
Sims C. and Zha T. (2006).
\textit{Were there regime switches in US monetary policy?}
American Economic Review 96, 54–81.

\bibitem[Stavroyiannis, Babalos, Bekiros and Lahmiri (2019)]{stavro}
Stavroyiannis S. , Babalos V. , Bekiros S. , and Lahmiri S. (2019).
\textit{The high frequency multifractal properties of Bitcoin}. Physica A: Statistical Mechanics and its Applications, 1873-2119, Vol. 520, p. 62-71.

\bibitem[Yermack (2014)]{yermack}
Yermack D. (2014).
\textit{Is Bitcoin a real currency? 
An Economic appraisal}.
New York: NYU - Stern School of Business.

\end{thebibliography}
\end{document}